\documentclass{acm_proc_article-sp}    

\usepackage{xspace}
\usepackage{graphicx}
\usepackage[pdftex,colorlinks=true,pdfstartview=FitV,
 linkcolor=black,citecolor=black,urlcolor=black]{hyperref}

\usepackage[normalem]{ulem} 
\usepackage{xcolor}

\usepackage{ifthen}
\usepackage{amssymb}
\newboolean{showcomments}
\setboolean{showcomments}{true} 
\ifthenelse{\boolean{showcomments}}
  {\newcommand{\Note}[3]{\fcolorbox{gray}{#3}{\bfseries\sffamily\scriptsize#1}{\sf\small$\blacktriangleright$\textit{#2}$\blacktriangleleft$}}}
  {\newcommand{\Note}[3]{}}

\newcommand{\Finding}[1]{\fbox{\begin{minipage}{0.98\linewidth}\textbf{Finding} \emph{#1}\end{minipage}}}

\newcommand{\x}{$\bullet$}
\newcommand\ie{i.e.,\xspace}
\newcommand\eg{e.g.,\xspace}
\newcommand\etal{\emph{et.al.}\xspace}
\newcommand\GM{General Motors\xspace}

\newcommand\insilico{in-silico\xspace}
\newcommand\MDE{model-driven engineering\xspace}
\newcommand\MBD{model-based design\xspace}

\title{Lessons Learned from Evaluating\\MDE Abstractions in an Industry Field Study}

\numberofauthors{3}

\author{
\alignauthor
Adrian Kuhn\\
       \affaddr{Software Practices Lab}\\
       \affaddr{Institute for Computer Science}\\
       \affaddr{University of British Columbia}\\
\alignauthor
Gail C. Murphy\\
       \affaddr{Software Practices Lab}\\
       \affaddr{Institute for Computer Science}\\
       \affaddr{University of British Columbia}\\
}

\begin{document}

\maketitle

\begin{abstract}

In a recent empirical study we found that evaluating abstractions of \MDE (MDE) is not as straight forward as it might seem. In this paper, we report on the challenges that we as researchers faced when we conducted the aforementioned field study. In our study we found that modeling happens within a complex ecosystem of different people working in different roles. An empirical evaluation should thus mind the ecosystem, that is, focus on both technical and human factors. In the following, we present and discuss five lessons learnt from our recent work.
\end{abstract}


\section{Introduction}

The promise of abstractions such as \MDE (MDE) is that the representations they provide to engineers are semantically more similar to the domain by hiding away implementation details. Yet, in a recent field study we found that identification and evaluation of these abstractions is not as straight forward as it might seem. Our work was an ethnographic study of MDE adoption from an engineer's perspective \cite{Kuhn2012x}, we did in-depth interviews with 20 engineers from the automative industry to learn how they succeed and struggle with model-driven techniques. While we present a brief overview of our study in \autoref{sec:study}, in this paper we focus on the challenges that we---as researchers---faced.

When doing our interviews we found it quite hard to capture the essence of abstractions. In the wild abstractions do not come as text book figures. We faced fundamental challenges such as, when are we looking at an abstraction and when not? When is an abstraction higher, \ie more abstract, than another? Does more abstraction offer more or less flexibility? What about abstractions that makes engineers more productive but are harder to understand? Or, an abstraction that is more understandable but, at times, leaky and thus leads to bursts of overhead work. Has it possibly been designed to solve another problem than what it is currently used for at this site of study? Or maybe even, are we looking at an organizational rather than a technical abstraction, \ie shift of responsibility between roles rather than between representation levels? Or, both? Essentially, how to identify an abstraction and how to evaluate the quality of that abstraction? 

Also we realized, engineers are not necessarily aware at which level of abstraction they are working. To them, what they do is their daily work. They might not even be aware of possible other abstraction layers below or above their work. So, asking them about abstraction levels often does not make sense in their language. For example, none of the engineers that we interviewed referred or thought of to their work as modeling, while in fact they were using model-driven engineering techniques.

As we conducted our field study we continuously improved our questionnaire as we gained more insight into the challenge of studying abstractions. In the following we summarize our lessons learned:

\vspace{-0.5\baselineskip}
\begin{itemize}
\item Quantitative approaches might fail, often abstractions are not ``more of the same'' but ``something different'' and thus not covered by established metrics.
\item While qualitative approaches are better suited, they might still miss parts of the picture unless the complete ecosystem is taken into account.
\item Recent adopters make for great interview partners, to them abstractions are still fresh and not yet daily business. Caution with learning curves is advised tough.
\item Asking for examples of most recent work items is para\-mount to avoid bias, or else a field study risks reporting prescribed processes rather than actual practices. 
\item Focus on communication patterns rather than artifact flow, as artifact flow is a subset of communication.
\end{itemize}
\vspace{-0.5\baselineskip}

These lessons learned are obviously based on a single case study only, however we hope that they might serve as the base for a discussion at the \textsc{Eesmod} workshop of how to best evaluate model-driven abstractions in field studies. 

The reminder of this paper is structured as follows:
In \autoref{sec:study} we discuss our choice of methodology, and provide a brief overview of our industry case study.
In \autoref{sec:meat} we discuss the lessons learned that are listed above.
In \autoref{sec:rel} we outline related work.
Eventually, in \autoref{sec:wrapup} we conclude with a summary.

\section{Our Case Study in a Nutshell}\label{sec:study}

Our research is aimed at understanding cognitive issues in \MDE, with a focus on information and communication needs from the perspective of individual engineers. As we started our research we had little background in \MDE and faced the choice of an appropriate research methodology. 

Quantitative approaches are warranted for scenarios where the state of the art provides a deep understanding of the domain under study and can refer to established means of measuring the phenomena under study. In our case, a literature review confirm a gap in research, we thus decided to go with an approach which is both qualitative and exploratory. 

To enable the gathering of detailed, rich and contextual information about model-driven engineering, we chose a qualitative study approach. We interviewed people working with \MDE technology in semi-structured interviews, following an exploratory case-study approach where open ended questions are asked in order to refine the research hypothesis as the current study is ongoing.

Exploratory approaches are great when you, as a researcher, start with little understanding of the domain under study. Grounded theory has recently gained popularity in the empirical software engineering community as an exploratory approach \cite{Steve2011a}. However, we choose to do an exploratory case study, where we are not necessarily looking for ``one theory'' but are facing a multifaceted case with possibly many theories explaining the observations.

An advantage of exploratory user studies, both grounded theory \cite{Corbin1990a} and exploratory case studies \cite{Yin2003}, is that, as a researcher, you are able to adapt your questionnaire as you learn from the participants answers. For example, as we started our user studies we had questions about ``how many models'' a person is responsible for. Though we quickly learned that, unlike to us academics, to the engineers ``model'' was not a countable term. The question did make as much as to them as asking how many ``softwares'' a software engineer in a traditional team is working on.

In our interviews, we found that the terms ``model'' and ``modeling'' were used ambiguously. Engineers generally did not refer to their work as ``modeling'' but used the terms ``auto-coding'' and ``hand-coding.'' These terms were used to differentiate between working with tools which include a step of code generation versus writing C-level code manually. Engineers used the term ``model'' ambiguously to refer to software models, as well as the \emph{plant models} used for the \insilico simulation of vehicles. Engineers also used the term ``simulation'' ambiguously to refer to running the \insilico simulation of the plant models, as well as to running software models from within the modeling tools as opposed to running the auto-generated sources. 

We believe the terminology we observed is mixing \MBD (MBD, an approach in system engineering for disentangling the development of control software and corresponding vehicles, using \insilico modeling while vehicles are not yet available) and \MDE (MDE). The ambiguous use of terminology can be explained if we look at \MDE as a division of labour between a few specialized language designers and many modelers. After all, the software engineers do not have to understand the full complexity of modeling, this is up to the specialized code-generation engineers. However, we found that points of friction in modeling tools, in particular the insufficient support of model diffing, may break the abstraction and nevertheless expose engineers to these complexities.

In the industry case study we interviewed 20 people working with model-driven approaches at General Motors, a large automotive company that heavily relies on \MDE for their software development. The study is going to appear in the proceedings of \textsc{Models 2012}, the hosting conference of this workshop\footnote{Preprint available at \url{http://arxiv.org/abs/1207.0855}}.

We found that, in the context of a large organization, contextual forces dominate the cognitive issues of using model-driven technology. The four forces we identified that are likely independent of the particular abstractions chosen as the basis of software development are the need for diffing in software product lines, the needs for problem-specific languages and types, the need for live modeling in exploratory activities, and the need for point-to-point traceability between artifacts. We also identified triggers of accidental complexity, which we refer to as points of friction introduced by languages and tools. Examples of the friction points identified are insufficient support for model diffing, point-to-point traceability, and model changes at runtime. 

In the following we are going to focus on the setup and setting of the case study as far as it is of interest for the present paper. The study consisted of interviews with 20 engineers and managers working in different roles. We visited the industry of interest (\GM) on two separate occasions, collecting data constructed through semi-structured in-depth interviewing. We interviewed 12 engineers and 8 managers. Overall, the engineers we interviewed came from four different teams from different company departments. All teams were global, that is spread across sites in India and America, however we interviewed people from the American sites only. The 12 engineers selected for interviews were sampled from several roles however their profiles are similar, that is they all work with the same process and use the same modeling technology. Each interview was 90--120 minutes long, recorded on tape and transcribed for encoding by one of the authors of this paper.

In a first visit, we interviewed 10 participants from both management and technical roles to familiarize ourselves with the software process used in the automotive industry. Based on what we learned from the first interviews, in our second visit, we interviewed an additional 10 participants, all of them working with software models but in different roles. The interviews were semi-structured,  following an exploratory case-study approach where open ended questions are asked in order to identify research hypothesis for  future studies. We asked participants to describe their work, how their work fits into the process of the organization, with whom they interact on a weekly basis, and which artifacts are the input and which are the output of their work. We also asked to see current or recent examples of artifacts on which they were working.

We transcribed the 12 interviews with engineers (4 from the first visit and 8 from the second visit). We encoded the transcripts and from this encoding, we distilled the contextual forces and points of friction presented in this paper. We encoded the interviews by tagging sentences with hashtags as if they were tweets. We then used a series of tag clouds to identify patterns in the data, merging and splitting tags as we saw need. We did two passes over the tags, a first one to identify all forces and frictions that shape the work of the participants, and a second pass to identify forces and frictions that might provide the basis for general hypotheses on model-driven engineering, ruling out those that are specific to the organization under study.

\section{Lessons Learned}\label{sec:meat}

In this section we share our lessons learnt of assessing abstractions in the wild. The nature of abstractions is difficult to qualify. There is value but often it is faceted, the same tool might be both more and less abstract. Sometimes providing better abstractions to one role but worse abstractions to another role, sometime to the same role. We found that, in the context of a large organization, contextual forces dominate the cognitive issues of using model-driven technology. While some abstractions are technical and applied locally by individuals, such as information hiding in programing language, MDE happens within a complex ecosystem of different people working in different roles, or even different parts of an organizations. 

In the following we present the main lessons learned from conducting our study.

\Finding{Quantitative approaches might fail, often abstractions are not ``more of the same'' but ``something different'' and thus not covered by established metrics.}

Novel abstractions, by their definition of raising representations to be more semantic and of hiding away details, are disrupting established quantitative metrics. So, how to establish good metrics for a novel abstraction? Designing new metrics requires a understanding of the domain, both technical and semantic. An understanding that is typically best gained by first doing qualitative and exploratory studies. 

It might seem obvious that KLOC (lines of code) are an inappropriate metric to compare the complexity of manually written and auto-generated code. Yet the difficulty of assessing abstractions using quantitative methods are much more subtle. While increasing the abstraction of one process step might positively impact that one step, work further downstream in the process might suffer negative benefits. For example, cost being gained by reduced development times might be outweigh by increased cost of certification, as has been found in a study by Hutchinson \etal\cite{Hutchinson2011a,Hutchinson2011b}. 

In a similar way, time gained by spending less time in one tool might be lost again by time spent in a novel tool or even doing unexpected ad-hoc workarounds. In our case study we have found that engineers spend significant time manually putting together screenshots of model changes in Microsoft Powerpoint in order to email them as a ``change set'' to other engineers. A job which has been automated using text-based diffing tools before the introduction of MDE abstractions. 

\Finding{While qualitative approaches are better suited (than quantitative ones), they might still be miss parts of the picture unless the complete ecosystem is taken into account.}

Participant will not report \emph{``see, part of my work is done by somebody else``} since to them it not their work anymore when somebody else does it. Yet, when looking at adoption of higher abstractions, such as MDE ,very often it is the case that part of somebody's work is now done by another role. Abstraction might move work from one role to another or even from one team to another. An empirical study of \MDE should thus mind the ecosystem and look into how details are abstracted away across roles and organizational units rather than just across technical boundaries. 


Sometimes, a comparison is being made between rhe shift from source to models in MDE and the shift from assembly to high-level languages. We found that this is not a useful analogy for designing empirical studies of MDE. The abstractions introduction when moving from assembly to high-level code are typically hiding away details which are taken care of by the machine. Consider for example garbage collection, where the machine takes care of allocating and freeing memory, which has previously been the engineer's burden. This is not the case of many of the abstractions introduced by MDE. Code generation is unlike code compilation. In order to provide engineers with domain-specific abstractions, the implementation details that are hidden from engineers must be taken care of by other, more specialized roles, rather than the machine. In our case study we found that a team of specialized code generation experts is responsible for defining and managing these abstractions. 

\Finding{Recent adopters make for great interview partners, to them abstractions are still fresh and not yet daily business. Caution with learning curves is advised tough.} 

Recent adopters are make for great interviews partners. They are much more aware of abstraction levels, since to them the abstractions their working with are novel and thus they are still aware of what improved and what worsened with introducing these novel abstractions. They were still in the state of comparing the Now (model driven engineering with Simulink and Rhapsody) and the Then (hand coding in machine level languages). Yet, caution is advised as they might be in a learning curve and some of their observations are due to that learning curve. But that is exactly our job as ethnographic researchers, taking a step back and being able to do this kind of analysis.

The very definition of abstraction means that we might prefer to ask certain kinds of questions about abstractions introduced to people who can compare how a similar things was built before, \ie they need to know what might be hidden now. More importantly, as we pointed out before, we need to ask how did people work before and what did the people versus the machine do compared to after the abstractions are introduced.

On our second visit we had the chance the include a team in our study who had only recently adopted \MDE. Interviewing this kind of recent adopters, offers a unique window of opportunity onto comparing the ``before'' and ``after'' of an abstraction's introduction. Other than engineers who had been using model-driven technology since their training, these people can compare how work has been done before and after the introduction of model-driven abstraction. 

A danger is though that these people might report negatively about many novel aspects of the abstractions as they are still on a learning curve of adopting new ways of getting work done. We found that it helped to have both recent adopters as well as experienced users of \MDE in the same sample.  

\Finding{Asking for examples of most recent work items is paramount to avoid bias, or else a field study risks reporting prescribed processes rather than actual practices.}

A quantum leap in our understanding of \MDE happened as we started to ask participants for concrete examples. This is a well-known lessons learned for many empirical researchers. If you, as a researcher, ask people general questions about how they get their job done, answers by participants tend to remain equally general, typically describing an idealized account of how their job should be done rather how their job is actual done when the ``rubber hits the road.''\footnote{\scriptsize \emph{This metaphor is just one of many wonderful examples of a rich language of automotive metaphors used by our interview partners.}}

Examples. Show us the most recent model (or other artifact in question) that you were working with. This focus on the most recent work is important to avoid the participant's filter bias. If not asked for the most recent artifact, or the most recent week, we risk of being presented with ideal-case examples rather than actual samples from their daily work. And they will look different than you as researcher expect! For example we asked engineers: do you have tests? Yes. Do you have repeatable tests? Yes. Are they automated? Yes. So we expected tests that are akin to unit testing practices known from software engineering. But when we asked them to show us their most recent test, they opened an Excel sheet with instructions to a human tester and with a fields to enter what they see on screen. Obviously repeated and automated, but not the way we expected!

For example, here is a lesson learned. After two visits of ten interviews each, we realized that we missed a whole population stakeholders, the code-generation experts. We interviewed product engineers and their leads but did reach out to that one specialized team of code-generation experts. We were neither aware that they exist nor which crucial role they play in the model-driven engineering process. Mainly because neither their job title gave any clue of their work, nor were they included in the process model that was presented to us. We only learned about them when we started to check in with wrap-up questions, e.g. ``Were there more people to whom you talked last week that are not yet on this list?,'' after having gone step by step through all communications and meetings of a participant's most recent week. So even in our participant's heads these specializes were so much outside of their process that they did not think of them in the first place. Yet, they are key to understanding the MDE adoption and its benefits. We thus plan to interview these people in future work.

We started to ask questions like ``tell us more of about the most recent change ticket your worked on'' and then continued with ``can you please show us the model that you changed for that ticket'' and then ``can you show us the exact changes you did to that model.'' This way we made sure to learn about the actual work being done, and not hearing twenty times about the same formally defined process that it has been presented to us on our first briefing.

It is very important to narrow down this kind of questions to the most specific examples possible, such as ``most recent change ticket you worked on'' or ``people you talked to last week,'' and then following them up with questions such as ``were there more people you talked to last week that had not been mentioned so far'' in order to make catch missing data. It is typically this latter kind of catch-up questions that reveal the most interesting tidbits of information.

Of course it may happen the that the ``most recent change ticket'' or ``last week'' had  been exceptional and do not to serve as good exemplars. We controlled for this using questions like ``had last week been a typical week of your work.'' Often the reply had been one of ``yes it had been typical expect for\dots'' which again lead to very interesting insight into their typical work as we learned which kind of work participants consider untypical. 

\Finding{Focus on communication patterns rather than artifact flow, as artifact flow is a subset of communication.}

In the begin our questionnaire focused on artifact flow, because we figured that would be a great source of learn about all parts of the ecosystem that are involved in MDE. But the we realized, focusing on people and communication is a much better approach: we started going with them through all meetings and communication of the most recent week, asking for each communication about every participant's role and how they are connected to each other. It was this information which proved to provide a complete picture of the MDE ecosystem, including artifact flow. We thus conclude that artifacts flow is a subset of communication only, and thus communication a better means to learn about abstractions and their adoption.

On our second visit we started to include questions such as ``whom did you talk to last week`` in our interviews. We asked participants to walk us through their most recent week of interaction with other people. This lead to most interesting insight about the organizational perspective of \MDE abstractions. For example we learned about teams and organization groups that had not been captured by the formalized process. For example, we learned that some engineers are part of task force groups that are evaluating novel \MDE technology. And, we learned about the existence of a specialized team of code generation experts of which we had not been aware of previously. Which has led to a revised understanding of \MDE not so much being a technical abstraction, where details are hidden away from humans and being taken care of by the machine (as \eg in a compiler), but an organizational abstraction where details are hidden away from a large workforce of engineers and being take care of by a specialized team of code generation experts \cite{Kuhn2012x}.

\section{Related Work}\label{sec:rel}

We are not the first to study MDE abstractions in the field, though to our best knowledge there is little work that shares lessons learnt about the research process used for conducting these studies. 

Heijstek and Chaudron \cite{Heijstek} studied an industrial MDE case over two years, where a team of 28 built a business application for the financial sector. Using grounded theory they found 14 factors which impact the architectural process. They found that MDE shifts responsibility from engineers to modelers, and that the domain-specific models facilitated easier communication across disciplines and even became a language of business experts. The setup of their case differs from our recent work~\cite{Kuhn2012x} in that their case had a whole-system view on a closed ecosystem of 28 people, with premium access to both project lead and main architect of the system. In comparison, we had a peephole view on a much larger ecosystem of tens of thousands of people that are collaborating across the main company and it's subsidiaries. It will be interesting to compare the findings of these two studies with regard to these different perspectives.

Hutchinson \etal presented their results of a qualitative user study, consisting of semi-structured interviews with 20 engineers in 20 different organizations \cite{Hutchinson2011a,Hutchinson2011b}. 
They identified lessons learned, in particular the importance of complex organizational, managerial and social factors, as opposed to simple technical factors, in the relative success, or failure, of MDE. As an example of organizational change management, the successful deployment of model driven engineering appears to require: a progressive and iterative approach; transparent organizational commitment and motivation; integration with existing organizational processes and a clear business focus.

The proceedings of RAO 2006 \cite{RAO2006}, a workshop on the role of abstractions, providing interesting insight into both the role and study of abstraction, both in the context of MDE and in the context of software engineering in general.


\section{Conclusions}\label{sec:wrapup}

In this paper we presented lessons learned with regard to the to research methodology of a recent industry case study of ours. 
When we reviewed definitions of \emph{abstraction} we found a common theme of ``hiding away details'' with the purpose of ``reducing details so programmers can focus on a few concepts at a time'' (quotes taken from Wikipedia). How can we, are we asking thus, evaluate what is hidden away?
We found that the introduction of MDE abstractions is not happening at the mere technical level, but happens within a complex ecosystem of different people working in different roles. An empirical study of MDE should thus mind the ecosystem. 

Given that abstraction is about hiding away details, often people working with these abstraction are not aware of what has been hidden themselves. We found that interviewing recent adopters of MDE technology provide a unique window of opportunity onto people who can compare how work has been done before and after an abstraction's introduction
We also found it helpful to ask participant to focus on concrete examples, such as ``most recent bug report'' or ``people you talked to last week,'' in order to learn about their actual work experience.  

These lessons learned are obviously based on a single case study only, however we hope that they might serve as the base for a discussion at the EESMOD workshop of how to best study ``that what is hidden away,'' \ie abstractions. 

\subsubsection*{Acknowledgments}

This research has been funded by the Canadian Network on Engineering Complex  Software Intensive Systems for Automotive Applications (NECSIS). Special thanks to Joe D'Am\-bro\-sia and Shige Wang from GM Research. We thank Deepak Azad, Neil Ernst, Manabu Kamimura for their feedback on an earlier draft of this paper.

\bibliographystyle{plain}
\bibliography{spl}

\end{document}